\documentclass[conference,10pt,a4paper]{IEEEtran}

\usepackage{silence}
\newlength{\flexwidth}
\usepackage{bm}
\usepackage{amsmath,amssymb,amsthm,fixmath}
\usepackage{mathtools}
\usepackage{accents}
\usepackage{color,colortbl}
\usepackage{xcolor}
\usepackage{siunitx}
\usepackage{cuted}
\usepackage{multirow,booktabs}
\usepackage{subcaption}
\usepackage[inline]{enumitem}
\usepackage{optidef}
\usepackage{graphicx}
\usepackage{lipsum}

\usepackage{amsmath}
\usepackage{epstopdf}
\usepackage[normalem]{ulem}
\newcommand{\revise}[2]{{\color{red}\sout{#1}}{\color{blue}#2}}

\usepackage[textsize=tiny,colorinlistoftodos]{todonotes}
\makeatletter
\define@key{todonotes}{bh}[]{
	\setkeys{todonotes}{author=\textbf{Bin}, color=lime!30}}%
\define@key{todonotes}{zf}[]{
	\setkeys{todonotes}{author=\textbf{Zexin}, color=blue!30}}%
\makeatother

\newif\ifreviewmode
\reviewmodetrue 
\reviewmodefalse

\ifreviewmode
\else
  \renewcommand{\todo}[1]{} 
  \renewcommand{\revise}[2]{#2} 

\usepackage[shortcuts,acronym,automake]{glossaries}
\makeglossaries
\newacronym{ue}{UE}{User Equipment}
\newacronym{bs}{BS}{base station}
\newacronym{csi}{CSI}{Channel state information}
\newacronym{cnn}{CNN}{Convolutional Neural Network}
\newacronym{fl}{FL}{Federated learning}
\newacronym{iot}{IoT}{Internet of Things}
\newacronym{mmwave}{mmWave}{millimeter-wave}
\newacronym{b5g}{B5G}{Beyond-Fifth-Generation}
\newacronym{6g}{6G}{Sixth Generation}
\newacronym{ml}{ML}{Machine learning}
\newacronym{sbs}{SBS}{small base station}
\newacronym{mu}{MU}{mobile user}
\newacronym{mbs}{MBS}{macro base station}
\newacronym{mse}{MSE}{Mean Squared Error}
\newacronym{cl}{CL}{centralized learning}
\newacronym{uav}{UAV}{unmanned aerial vehicle}
\newacronym{bme}{BME}{Bayesian Model Ensemble}
\newacronym{iid}{IID}{independent and identically distributed}
\newacronym{raf}{RAF}{robust aggregation function}
\newacronym{sgd}{SGD}{stochastic gradient descend}
\newacronym{cdf}{CDF}{cumulative distribution function}
\newacronym{lid}{LID}{local intrinsic dimensionality}
\newacronym{llpf}{LLPF}{local loss pre-filtering}
\newacronym{mitm}{MITM}{man-in-the-middle}
\newacronym{ae}{AE}{adversary entitie}
\newacronym{tof}{TOF}{time of fly}
\newacronym{rss}{RSS}{received signal strength}
\newacronym{jsr}{JSR}{jamming to signal ratio}
\newacronym{3d}{3D}{three dimensional}
\newacronym{aoa}{DoA}{direction of arrival}
\newacronym{sdp}{SDP}{semi-definite programming}
\newacronym{jss}{JSS}{jamming signal strength}
\newacronym{pdr}{PDR}{packet delivery ratio}
\newacronym{wcl}{WCL}{weighted centroid localization}
\newacronym{asl}{ASL}{adaptive least-squares}
\newacronym{nlos}{NLOS}{Non-Line-of-Sight}
\newacronym{snr}{SNR}{signal to noise ratio}
\newacronym{wvgd}{WVGD}{weighted vector-based gradient descend}
\newacronym{esprit}{ESPRIT}{estimation of signal parameters via rotational invariance technique}
\newacronym{music}{MUSIC}{multiple signal classification}
\newacronym{fos}{FOS}{fast orthogonal search}
\newacronym{crb}{CRB}{Cramer-Rao bound}
\newacronym{lse}{LSE}{least squared estimation}
\newacronym{wlse}{WLSE}{weighted least squared estimation}
\newacronym{gd}{GD}{Gradient descend}
\newacronym{rgd}{RGD}{robust gradient descend}
\newacronym{mle}{MLE}{maximium likehood estimation}
\newacronym{spgd}{SPGD}{sample pruning gradient descend}

\usepackage{tcolorbox}
\tcbuselibrary{many}

\newtcbtheorem[]{alg}{Algorithm}{fonttitle=\bfseries}{alg}

\usepackage[norelsize,linesnumbered,ruled]{algorithm2e}
\SetKwRepeat{Do}{do}{while}
\makeatletter
\newcommand{\removelatexerror} {\let\@latex@error\@gobble}
\makeatother

\begin{document}
	
	\title{A Robust UAV-Based Approach for Power-Modulated Jammer Localization Using DoA}
	
	\author{
		\IEEEauthorblockN{Zexin~Fang\IEEEauthorrefmark{1},~Bin~Han\IEEEauthorrefmark{1},~and~Hans~D.~Schotten\IEEEauthorrefmark{1}\IEEEauthorrefmark{2}}
		\IEEEauthorblockA{
  \IEEEauthorrefmark{1}{University of Kaiserslautern (RPTU), Germany}\\\IEEEauthorrefmark{2}{German Research Center for Artificial Intelligence (DFKI), Germany}
		}
	}
	
	\bstctlcite{IEEEexample:BSTcontrol}
	
	\maketitle

	\begin{abstract}  
       \Glspl{uav} are well-suited to localize jammers, particularly when jammers \revise{situated in}{are at} non-terrestrial locations, where conventional detection methods face challenges. \revise{Considering the potential for multiple power-modulated jammers, which can undermine the effectiveness of many localization algorithms, our proposed method, \gls{spgd}, offers robust localization with low computational complexity.}{In this work we propose a novel localization method, \gls{spgd}, which offers robust performance against multiple power-modulated jammers with low computational complexity.}
		
	\end{abstract}
	
	
	\IEEEpeerreviewmaketitle
	
	\glsresetall

	\section{Introduction}\label{sec:introduction}
    In recent years, \glspl{uav}, especially low-altitude \glspl{uav}, are becoming a popular solution for reliable wireless communication in a variety of applications \cite{TJGJsecureUAV2019}. Compared to current terrestrial base stations, \gls{uav}-mounted base stations can surpass terrain-related signal propagation issues, while also offering solutions for unexpected and temporary communication demands \cite{ZYGMsecureUAV2018}. Conversely, \gls{uav}-mounted jamming attacks can be more effective against terrestrial communication systems compared to terrestrial jammers \cite{WQWRjamminguav2019}. Considering that a hovering \gls{uav} is easily noticeable, a sophisticated attacker may place \gls{uav}-mounted jammers or drop jammers with \gls{uav} discreetly in inconspicuous locations, such as bird nests or building rooftops. Additionally, those jammers can also be in disguise to blend in with the environment, making them harder to spot. Given the potential risk outlined above, \revise{it's}{it is} crucial to locate jammers in a \gls{3d} space. Assuming the jammer is omni-directional, passive jammer localization methods such as \gls{pdr} \cite{QXJTpdr2015}, \revise{or }{}\gls{wcl} \cite{WXWSurvJLoc2016}\revise{}{,} and \gls{asl} method \cite{LZHWYLSQ2012}\revise{,}{} can be used for jammer localization. However, those methods commonly require a massive number of affected network nodes and \revise{can't}{cannot} provide altitude estimation.
    
    In the scope of high-accuracy \gls{3d} source localization in wireless networks, joint \gls{rss} or \gls{aoa} measurements can be used to localize source with linear estimator \cite{HAAjRSSD2020}. Hybrid \gls{rss} and \gls{aoa} measurements can also be used to precisely estimate source localization with \gls{sdp} algorithms \cite{QHLXsourceloc2020}. In the mean time, \glspl{uav} are also considered well-suited for \gls{aoa} measurements because they can operate with minimal \gls{nlos} interference. This feature ensures high-accuracy \gls{aoa} estimation, enabling precise source localization with \glspl{uav} \cite{SBYAOAloc2023}.

    \revise{Given that}{Practically,} the transmitting power and antenna patterns of jamming sources are usually unknown\revise{, and in a worst-case scenario}{. In some scenarios}, jammers \revise{might be}{can be even} power-modulated, constantly altering their transmission power\revise{,}{. In this case,} localization methods based on \gls{rss} can be \revise{highly }{}unreliable, as evidenced in \cite{TDGjammingpow2022}\revise{. This is due to that \gls{rss}-based localization relies}{, due to their dependency} on the a constant transmission power to estimate \revise{}{the} distance. On the other hand, \gls{aoa}-based localization \revise{doesn't}{does not} require the transmission power to be constant. Instead, \revise{\gls{aoa} measurement}{their} accuracy is linked with \gls{jsr}, \revise{}{as }revealed in \cite{OAMjammDoA2020}. This suggests that \gls{aoa}-based localization tends to be more resilient than \gls{rss}-based localization, when transmission power and antenna patterns are unknown.

   Most studies on \gls{aoa} localization assume a fixed normal distribution for the estimation error, \revise{highlighting the need for more sophisticated modeling}{reveals a gap in accounting for the variability and complexity of real-world jammer behaviors and radio environment}. In \revise{our}{this} work, we consider directional jammer antenna patterns and \gls{aoa} measurement errors dependent on the \gls{jsr}. Unlike prior research, we acknowledge the uncertainty of \gls{aoa} resolution in \gls{uav}-based jammer localization due to power modulation of jammers. To address this, we propose \gls{spgd} for robust localization in such scenarios. The \revise{}{remainder of this} paper is structured as follows: In Sec.~\ref{sec:sys_model}, we introduce the propagation model and discuss \gls{aoa} estimation error. Sec.~\ref{sec:loctech} presents localization techniques using \gls{aoa}, while Sec.~\ref{sec:eva} validates these techniques through numerical simulations. Sec.~\ref{sec:con} concludes this work.

	\section{Preliminary}\label{sec:sys_model}
    \subsection{Propagation model}
   Jamming signals attenuate as they travel through the air, influenced by environmental factors like obstacles, terrain, and weather conditions. As a result, the received jamming power can be modeled by a log-normal path loss model \cite{TDGjammingpow2022}:
    \begin{equation}\label{eq:pathloss}
		P_\mathrm{j}\left(d\right)= P_\mathrm{j}\left(d_0\right) - 10n_\mathrm{p} \mathrm{log}\left(\frac{d}{d_0}\right)+ X_{\sigma},
	\end{equation}
    \begin{equation}\label{eq:pathlossN}
		X_{\sigma}\sim\mathcal{N}\left(0,\sigma\right),
	\end{equation}
   where $P_\mathrm{j}\left(d\right)$ represents the received jamming signal power at distance $d$, and $P_\mathrm{j}\left(d_0\right)$ denotes the jamming power at the reference distance $d_0$. $n_p$ signifies the path loss factors. Additionally, $X{\sigma}$ is a log-normal random variable with a mean and standard deviation, thereby modeling the environmental factors. The \gls{jsr} can then be defined as, 
   \begin{equation}\label{eq:jsr}
		\mathrm{JSR} = P_{\mathrm{j}}\left(d_j\right) - P_{\mathrm{s}}\left(d_s\right), 
	\end{equation}
   where $d_j$ represents the distance from the jammer to the receiver, and $d_s$ indicates the distance from the signal source to the receiver.
   
   \subsection{\gls{aoa} estimation model}
   \gls{aoa} estimation is a well-established field with various techniques, including \gls{music}, \gls{esprit}, and \gls{fos} \cite{MUSI1986,FOSDOAKM1988,ESDOARRK1989}. 
   \gls{aoa} estimation involve two parameters: elevation and azimuth. Elevation refers to the angle in the vertical plane, measuring the height relative to the horizontal plane. Azimuth, on the other hand, is the angle in the horizontal plane.
   
\begin{figure}[!htbp]
		\centering
		\includegraphics[width=0.82\linewidth]{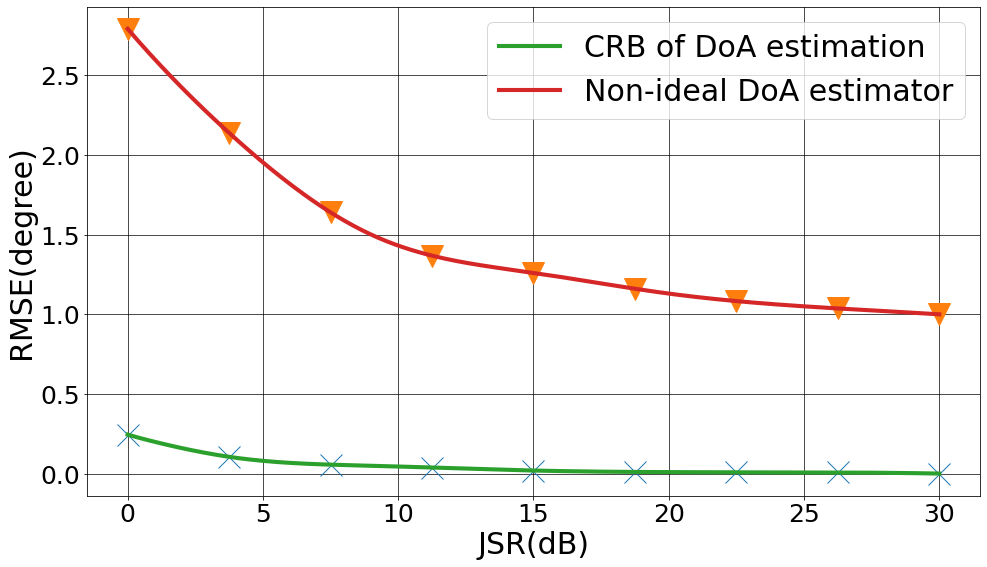}
		\caption{CRB of \gls{aoa} estimation and non-ideal \gls{aoa} estimator}
		\label{fig:crbDoA}
	\end{figure}

   Considering $P_j \gg P_s$ in jamming affected area, the signal can be seen as interference to jamming \gls{aoa} estimation. In \cite{OAMjammDoA2020}, it is shown that the \gls{crb} is associated with the \gls{jsr}. The finding also indicated that the performance of any non-ideal \gls{aoa} estimator will converge above the \gls{crb}, depicted in Fig.~\ref{fig:crbDoA}. Thus, the noisy \gls{aoa} estimation $\left[\theta_n^{\circ},\phi_n^{\circ}\right]^{T}$ can be modelled as,
   \begin{equation}\label{eq:azimu}
	      \theta_n^{\circ}= \theta_n + e_{\mathrm{d}},
   \end{equation}
   \begin{equation}\label{eq:ele}
	      \phi_n^{\circ}= \phi_n + e_{\mathrm{d}},
   \end{equation}
    \begin{equation}\label{eq:errornoa}
		e_{\mathrm{d}}\sim\mathcal{N}\left(0,\frac{\sqrt{\sigma_{\mathrm{d}}}}{2}\right),
	\end{equation}
   where $\theta_n\in\left(-\pi,\pi\right)$ and $\phi_n\in\left(-\frac{\pi}{2},\frac{\pi}{2}\right)$ are the true value of azimuth and elevation, respectively; $\sigma_{\mathrm{d}}$ indicates the error power of \gls{aoa}. 
   
    \section{Methodology}\label{sec:loctech}
    \subsection{Jammer localization via \gls{lse}}\label{subsec:LSE}
    The jammer localization problem can be formulated as an optimization problem, which can be solved with \gls{lse}. This optimization problem has been formulated \cite{LSEW2021},
    \begin{equation}\label{eq:LS}
		\hat{\mathbf{p}} = \underset{\mathbf{p}}{\operatorname{argmin}}\sum_{n=1}^N\bigg\{\big(\mathbf{o}_{1n}^T\left(\mathbf{p} - \mathbf{p}_n^\circ\right)\big)^2+\big(\mathbf{o}_{2n}^T(\mathbf{p} - \mathbf{p}_n^\circ)\big)^2 \bigg\}
	\end{equation}
    where $\mathbf{o}_{1n}$ and $\mathbf{o}_{2n}$ are the orthogonal vectors, with
    \begin{equation}\label{eq:OV1}
		\mathbf{o}_{1n} = \left[-\sin{\theta_n}, \cos{\theta_n}, 0\right]^T,
	\end{equation}
    \begin{equation}\label{eq:OV2}
		\mathbf{o}_{2n} = \left[\cos{\theta_n}\sin{\phi_n},\sin{\theta_n}\sin{\phi_n}, -\cos{\phi_n}\right]^T. 
	\end{equation}
    Meanwhile, $\mathbf{p}_n^\circ$ indicates position of \gls{uav} with estimation error, considering $\mathrm{p}_n^\circ = \left[x_n + e_{\mathrm{p}},y_n+e_{\mathrm{p}},z_n+e_{\mathrm{p}}\right]^T$ and $e_{\mathrm{p}}\sim \mathcal{N}\left(0,\frac{\sqrt{\sigma_\mathrm{p}}}{3}\right)$. Similarly, $\sigma_\mathrm{p}$ is position error power. 

    The closed-form solution of Eq.(\ref{eq:LS}) can be obtained, 
    \begin{equation}\label{eq:LSEsolu}
		\hat{\mathbf{p}} = (\mathbf{A}^T\mathbf{A})^{-1}\mathbf{A}^T\mathbf{b},
	\end{equation}
    where 
    \[
     \mathbf{A} = \begin{bmatrix}
    \mathbf{o}_{11}^T  \\
    \vdots  \\
    \mathbf{o}_{1N}^T\\
    \mathbf{o}_{21}^T  \\
    \vdots  \\
    \mathbf{o}_{2N}^T\\
    \end{bmatrix};
    \mathbf{b} = \begin{bmatrix}
     \mathbf{o}_{11}^T\mathbf{p}_1^\circ  \\
    \vdots  \\
    \mathbf{o}_{1N}^T\mathbf{p}_N\circ\\
    \mathbf{o}_{21}^T\mathbf{p}_1\circ  \\
    \vdots  \\
    \mathbf{o}_{2N}^T\mathbf{p}_N\circ\\
    \end{bmatrix}.
    \]
    
    In many cases, the performance of \gls{lse} can be further improved by incorporating a weight matrix to differentiate estimations. In \cite{LSEW2021}, a weight matrix is constructed with respect to the distance to the source with $w_n = 1 - \frac{d_n}{\sum_{n=1}^N d_n}$. However, in our scenario where the antenna pattern and transmitting power of the jamming source are unknown, modeling the weight with the distance to the source is not feasible. Alternatively, considering that the measurement error is linked with $\mathrm{JSR}$, it can be utilized to construct a weight matrix. For $\mathbf{J} = \left[\mathrm{JSR}_1,\dots, \mathrm{JSR}_n, \mathrm{JSR}_1,\dots,\mathrm{JSR}_n\right]^T$, a weight matrix can be formulated,
    \begin{equation}\label{eq:weight}
		\mathbf{W} = \frac{2\times 10^{\frac{\mathbf{J}}{2n_\mathrm{p}}}}{\sum\left(\mathbf{J}\right)}.
	\end{equation}
    Then, the closed-form solution of \gls{wlse} can be written,
     \begin{equation}\label{eq:WLSEsolu}
		\hat{\mathbf{p}} = \left(\mathbf{A}^T\mathbf{W}\mathbf{A}\right)^{-1}\mathbf{A}^T\mathbf{W}\mathbf{b}.
	\end{equation}

   As many of previously mentioned studies have shown, \gls{lse} and \gls{wlse} can achieve a localization accuracy within a few meters when $\sigma_\mathrm{d}$ is several degrees. While more precise techniques like \gls{mle} exist, they typically require much higher computational resources. Considering that jammer localization within several meters is sufficient, \gls{lse} and \gls{wlse} offer a practical balance between precision and efficiency.  
    \subsection{\gls{spgd} method}
\begin{figure}[!htbp]
		\centering
		\includegraphics[width=0.65\linewidth]{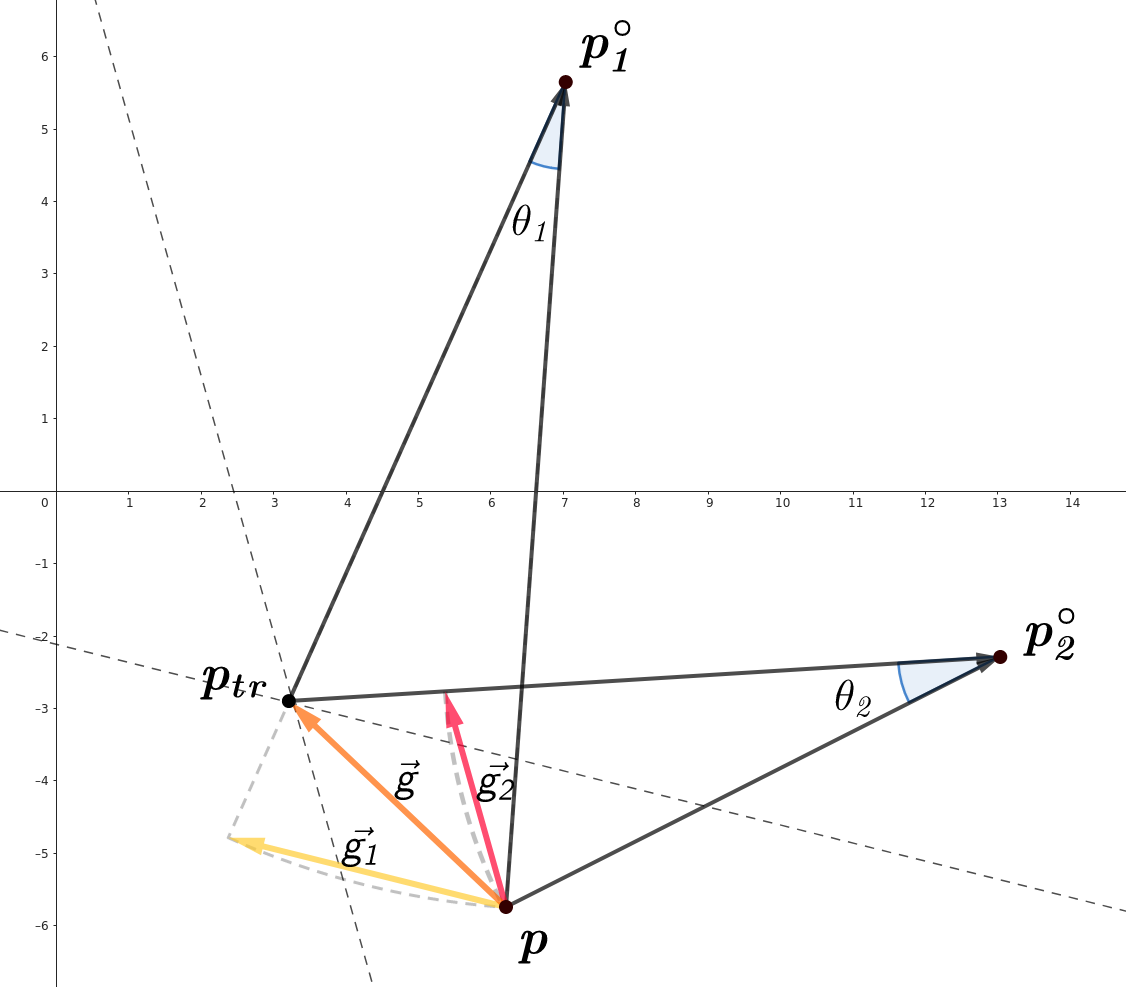}
		\caption{Gradient vectors of \glspl{aoa}}
		\label{fig:DiaUAV}
	\end{figure}

    The \gls{gd} algorithm is commonly used to solve localization problems \cite{GDzbs2023}. It is highly valued for its adaptability and computational efficiency. The optimization problem can be rewritten as minimizing the gradient vectors (depicted in Fig.~\ref{fig:DiaUAV}), 
     \begin{equation}\label{eq:GDoptima}
		\hat{\mathbf{p}} = \underset{\bm{p}}{\operatorname{argmin}}\sum_{n=1}^N\mathbf{g}_n,
	\end{equation}
    \begin{equation}\label{eq:GradientVector}
    \mathbf{g}_n = - \left(\mathbf{p} - \mathbf{p}_n^\circ + \mathbf{u}_n \times||\mathbf{p}-\mathbf{p}_n^\circ||\right) ,
	\end{equation}
    where $\mathbf{u}_n$ is the unit vector from the true position $\mathbf{p}_\mathrm{tr}$, 
     \begin{equation}\label{eq:unitvector}
         \mathbf{u}_{n}= \left[\frac{1}{\sec\theta_n^\circ \sec\phi_n^\circ}, 
          \frac{\sin\theta_n^\circ}{\sec\phi_n^\circ}, \sin\phi_n^\circ\right]^T.
	\end{equation}
 
    To ensure robust estimation while reducing computational complexity, we propose the \gls{spgd} method. At each iteration, a proportion of samples with the largest errors relative to the current estimation are discarded. The detailed algorithm is outlined in Algorithm~\ref{alg:spgd}.  
    \begin{algorithm}[!htbp]
    \caption{\gls{spgd}}
    \label{alg:spgd}
    \scriptsize
    \DontPrintSemicolon
    Input: A set contains all coordinate $\mathcal{P}$ ; a set contains all unit vectors $\mathcal{U}$; learning rate and learning decay factor  $\alpha$ and $\beta$; sample pruning rate $\eta$; iteration number $K$\\
    Output: estimated coordinate $\mathbf{p}$. \\
    \SetKwProg{Fn}{Function}{ :}{end}
    \Fn{\emph{SPGD}}{ 
        get $\mathcal{P}$ and $\mathcal{U}$; \\
        initialize $\mathbf{p} = \frac{1}{N}\sum^N_{n=1}\mathbf{p}_n^\circ$; \\
        \For {$k = 1:K$ }
        {$\mathbf{g} = \sum^N_{n=1}\mathbf{g}_n $;\\
        $\mathbf{p} = \mathbf{p} + \alpha\times\mathbf{g} $;\\
        $\alpha = \alpha\times\beta $; //\text{reduce learning rate}\\
        $\mathcal{D} \leftarrow \big\{d_n = \big\Vert \frac{\mathbf{p}-\mathbf{p}_n^\circ}{\Vert\mathbf{p}-\mathbf{p}_n^\circ\Vert}\big\Vert-\mathbf{u}_n ; 1 \leq n \leq N \big\}$; //\text{errors to current estimation}\\ 
        $n_\mathrm{r} = \max\left(N\times\eta,1\right)$;\\
        $\mathcal{N}_r\leftarrow\{ j =\underset{n\leq n_r}{\operatorname{argmax}} ; d_n \}$ ; //\text{select samples with largest errors}\\
        \If{$N - n_\mathrm{r}\geq 3$}{\For{$j\in\mathcal{N}_r$}{remove $j_\mathrm{th}$ elements from $\mathcal{P}, \mathcal{U}$}} 
        $N = N-n_r$
        }
        Output $\mathbf{p}$}
    \end{algorithm}
   
    \section{Evaluation of Methodology}\label{sec:eva}
    \subsection{Jammer antenna pattern}\label{subsec:jammerradi}
\begin{figure}[!htbp]
		\centering
		\includegraphics[width=0.75\linewidth]{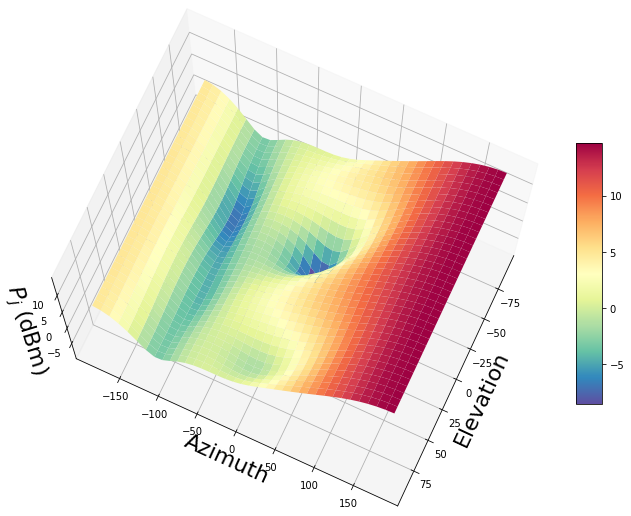}
		\caption{Antenna pattern of helical antenna}
		\label{fig:radiationjammer}
	\end{figure}

    In this work, we model the jammer's antenna pattern as directional, utilizing a helical antenna. Its main lobe is oriented towards $\theta = \pi$, as depicted in Fig.~\ref{fig:radiationjammer}. The helical antenna exhibits a dynamic range of $20\si{\dB}$.
    \subsection{Jammer Localization in an ideal scenario}
    To assess the localization performance of the aforementioned techniques, we consider an ideal scenario where the \gls{uav} cruises within a large area. The jammer $A$ is positioned at the center with a certain level of ambiguity. We assume the signal source is located very far away, with $P_j \gg P_s$. 
    \begin{table}[!htbp]
		\centering
        \scriptsize
		\caption{Simulation setup 1}
		\label{tab:setup1}
		\begin{tabular}{>{}m{0.15cm} | m{0.9cm} l m{3.0cm}}
			\toprule[2px]
			&\textbf{Parameter}&\textbf{Value}&\textbf{Remark}\\
			\midrule[1px]
            &$ $& {\scriptsize$[0\sim100, 0\sim100, 5\sim25]$}& Cruising area\\    
			  &$ $& {\scriptsize$[40\sim60, 40\sim60, 12\sim18]$}& Jammer location area\\
			\multirow{-1.5}{*}{\rotatebox{90}{\textbf{System}}} &$\max(Ptj)$&$5\sim25 \si{dBm}$& jammer main lob transmitting power \\
            &$Ps$&$-15\si{dBm}$& received signal power\\
            &$N$&$8,20,40$& number of \glspl{aoa} measurements\\
            &$\sigma_\mathrm{P}$&$3$& position error power\\
			    \midrule[1px]
            &$K$ & $10$& Iteration number\\
            &$\alpha$ & $1$& learning rate\\
			&$\beta$ & $0.7$& learning decay factor\\
	          \multirow{-3.8}{*}{\rotatebox{90}{\textbf{SPGD}}}&$\eta$ & $0.3$& Sample pruning rate\\
            \bottomrule[2px]
		\end{tabular}
	\end{table}

\begin{figure}[!htbp]
		\centering
		\includegraphics[width=0.88\linewidth]{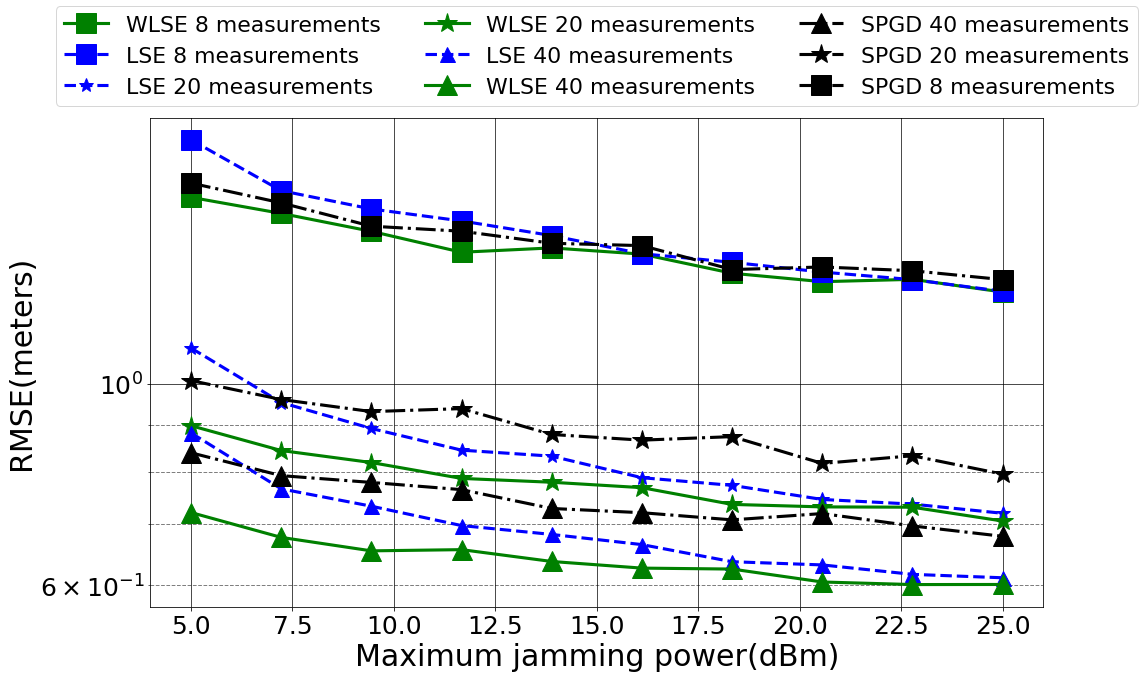}
		\caption{Antenna pattern of helical antenna}
		\label{fig:PerformIdeal}
	\end{figure}

    All simulation results in this paper, including those depicted in  Fig.~\ref{fig:PerformIdeal} under an ideal scenario, are obtained from $2000$ Monte Carlo iterations. \gls{wlse} exhibited the best performance across three difference number of $N$, especially when $\max(P_{tj})$ is low. \gls{spgd} demonstrated a similar performance to \gls{wlse} under $8$ estimations are collected. Meanwhile \gls{spgd} also shows better performance than \gls{lse} while $\max(P_{tj}) = 5$. This indicates a resilience of \gls{spgd} to estimation error, especially under limited estimations are collected, where estimation errors are not likely to cancel out. In other cases, \gls{lse} outperformed \gls{spgd}. However, \gls{lse} requires multiple matrix multiplications, leading to computational complexity that is predominantly $\mathcal{O}(2Nq^2)$, where $q$ is the number of sample features, and in this context, $q=3$.
    \gls{wlse} is more than $\mathcal{O}(2Nq^2)$.  Since $N$ is reduced exponentially with \gls{spgd}, the computational complexity of \gls{spgd} is predominately $\mathcal{O}\left(q\frac{N(1-\eta^K)}{1-\eta}\right)$. In scenarios where $N \gg K$, \gls{spgd} demonstrates significantly lower computational complexity compared to \gls{lse}. As discussed earlier in Subsec.~\ref{subsec:LSE}, \gls{spgd} achieves a balanced trade-off between accuracy and computational complexity.
    \subsection{Jammer localization in a non-ideal scenario}
    The presence of power-modulated jammers and other strong signal sources can significantly complicate the localization problem. While many \gls{aoa} estimation algorithms can resolve multiple \glspl{aoa} at an increased computational cost, they face significant challenges when dealing with highly coherent signals, such as jamming signals. The author in \cite{OAMjammDoA2020} demonstrated that the \gls{aoa} of two sinusoidal jamming signals with different frequencies can be resolved when  $P_{jA} = P_{jB}$. However, this is only theoretically possible. In practice, due to differences in distance, antenna directionality, and other environmental factors, the condition $\vert P_{jA}  - P_{jB}\vert \gg 0$ is more likely to hold. In this scenario, accurately resolving the \gls{aoa} of jamming components other than the largest is unlikely. In contrast to static receiver settings, where the \glspl{aoa} of multiple sources can be spatially clustered, associating these signal \glspl{aoa} with the correct sources presents a significant challenge when a cruising \gls{uav} serves as the receiver.

    To simplify localization and minimize computational costs, we consider only the \gls{aoa} of the largest jamming component is solvable, and other components as interference. Assuming $M$ jammers are involved, and given that the cruising center of the \gls{uav} is randomly determined with very limited information, the probability of it coinciding with the geometric center is very low. Combining with other random factors, the equal probabilities of estimation $\mathrm{P}_A = \mathrm{P}_B = \dots = \mathrm{P}_M$ can be ignored. Thus, the jammer with the largest probability $\mathrm{P}$ can be expected to be localized with minimal efforts. We conduct simulations to localize jammer $A$ with \(M = 2, 3\) and \(\mathrm{P}_A \in \left[\frac{1}{M}, 1\right]\), detailed setup follow Tab.~\ref{tab:setup1} with $N = 40$. The cruising center's lean toward a jammer will also affect localization, so we consider two scenarios: one with a strong lean toward jammer $A$, and another with a slight lean toward jammer $A$.

\begin{figure}[!htpb]
		\centering
		\begin{subfigure}{.85\linewidth}
			\centering
            
			\includegraphics[width=\linewidth]{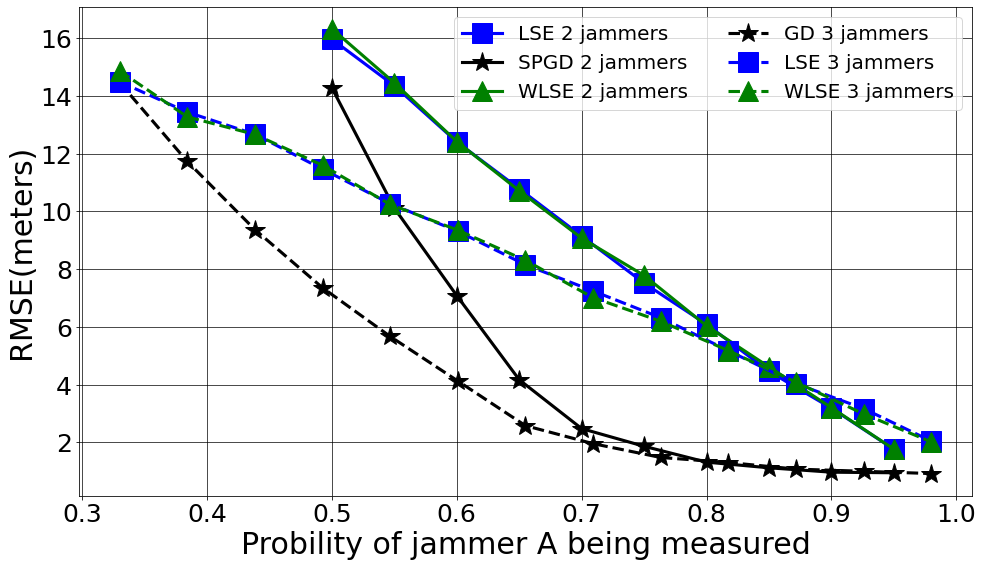}
			\subcaption{Case 1: strongly leans to jammer $A$}
			\label{subfig:gkloc}
		\end{subfigure}
		\begin{subfigure}{.85\linewidth}
			\centering
			\includegraphics[width=\linewidth]{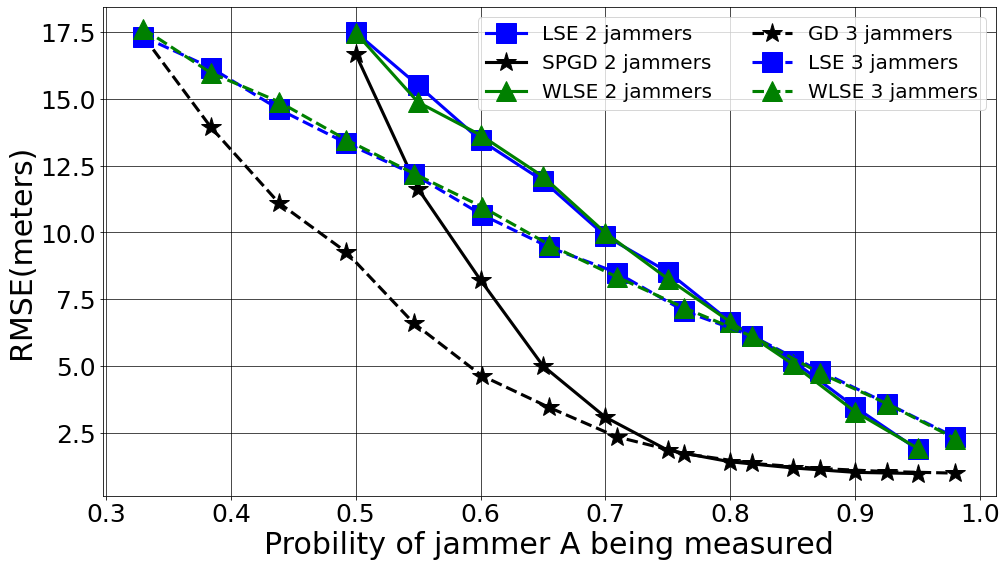}
			\subcaption{Case 2: slightly leans to jammer $A$}
			\label{subfig:bkloc}
		\end{subfigure}
     \caption{Localization performance under varying $P_A$}\label{fig:locprob}
	\end{figure}

    The simulation results are shown in Fig.~\ref{fig:locprob}. In both cases, the localization error of \gls{spgd} decreased exponentially as $\mathrm{P}$ increased, while the errors for \gls{wlse} and \gls{lse} decreased linearly. \gls{spgd} was able to provide much more robust estimations in complex scenarios by leveraging the correlations among estimations. Meanwhile, a slight degradation in the performance of all localization techniques in \emph{case 2} can be observed. This occurs because \gls{aoa} measurements from other sources in \emph{case 2} tend to lean more toward a direction orthogonal to that of jammer $A$, resulting in greater localization errors.
	  \begin{figure}[!htpb]
		\centering
		\begin{subfigure}{.85\linewidth}
			\centering
            
			\includegraphics[width=\linewidth]{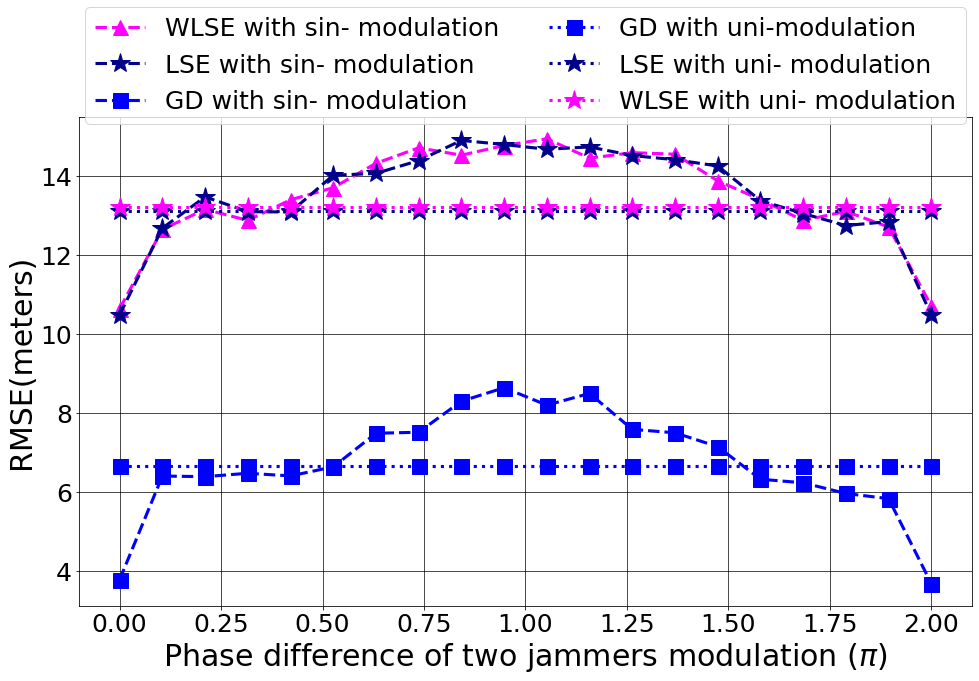}
			\subcaption{localization errors with regard to modulation phase difference}
			\label{subfig:phaseRMSE}
		\end{subfigure}
		\begin{subfigure}{.85\linewidth}
			\centering
			\includegraphics[width=\linewidth]{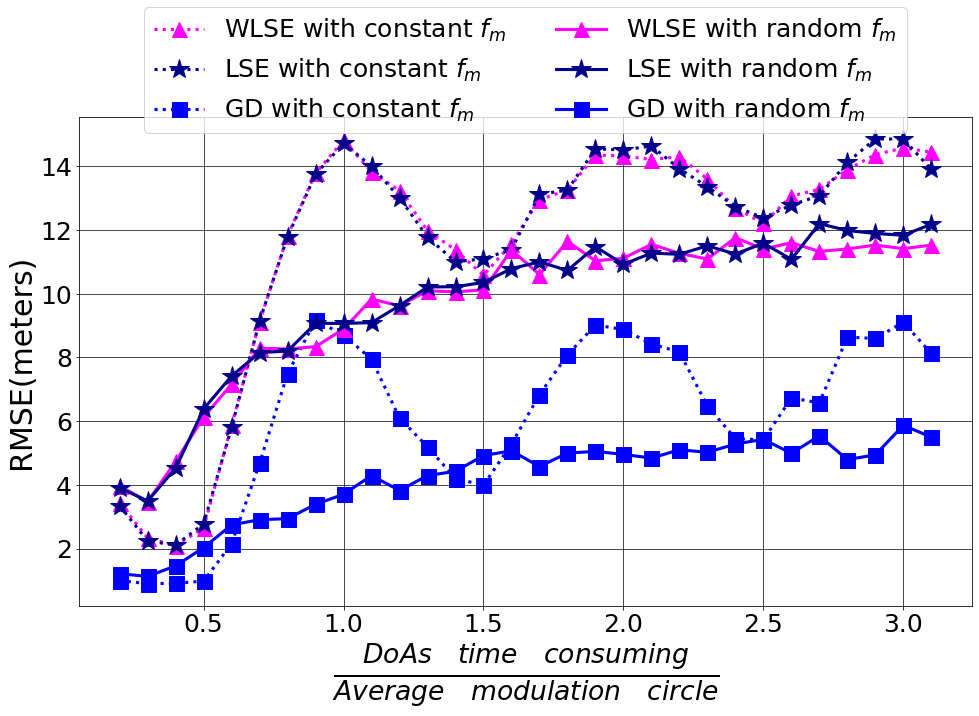}
			\subcaption{localization errors in relation to time consuming}
			\label{subfig:cirleRMSE}
		\end{subfigure}
     \caption{localization errors under different modulation schemes}\label{fig:locprow}
	\end{figure}
     \subsection{Jammer localization under different power-modulation schemes}
     In this subsection, we delve into the impact of various jammer power-modulation schemes on localization performance. Specifically, we consider the arrangement of two jammers. The jammer was described in Subsec.~\ref{subsec:jammerradi}, with main lobe transmitting power $\max(P_{tj})$ falling between $5 \si{dBm}$ and $25 \si{dBm}$. Initially, we consider two power-modulation schemes:
     \begin{enumerate}[label=\emph{\roman*)}]
        \item \textbf{random uniform modulation}: $\max\left(P_{tj}^A\right)\sim\mathcal{U}(5,20)$; $\max(P_{tj}^B)\sim\mathcal{U}(5,20)$
        \item \textbf{sinusoidal modulation}: $\max\left(P_{tj}^A\right) = 12.5 + 7.5\sin{\frac{t}{T}}$; $\max\left(P_{tj}^B\right) = 12.5 + 7.5\sin{\left(\frac{t}{T}+\Phi\right)}$. 
    \end{enumerate}
     We consider \gls{aoa} estimations start randomly and the time consuming $T_\mathrm{m}$ equals to the modulation circle $T$. 
     The simulation results are shown in Fig.~\ref{subfig:phaseRMSE}. 
    
    Notably, \gls{spgd} outperformed \gls{wlse} and \gls{lse} across all power-modulation schemes.
    When the phase difference $\Phi = 0$, it indicates that $\max\left(P_{tj}^A\right)$ and $\max(P_{tj}^B)$ vary equally. This uniformity, coupled with the tendency of the randomly determined cruising center to lean towards one jammer, leads to most resolved \glspl{aoa} being attributed to that jammer. Therefore, at $\Phi = 0$, the impact of \textbf{sinusoidal modulation} on localization performance is only marginal. At $\Phi = \pi$, the absolute power differences $\left\vert\max(P_{tj}^B) - \max\left(P_{tj}^A\right)\right\vert$ reaches its maximum. Consequently, even if the cruising center lean towards one jammer, the other jammer can still significantly interfere with localization, resulting in large errors.
    
    In \textbf{sinusoidal modulation}, the frequency $f_m = \frac{1}{T}$ can either remain constant or exhibit some degree of randomness, represented by $f_m = b\overline{f}_m$ and $ b\sim\mathcal{U}\left(0, 2\right)$. When $\Phi = \pi$, sinusoidal modulation can significantly affect localization; however, while \gls{aoa} estimations are done in a short time can mitigate this effect. We demonstrate this relationship with $T_m$ ranging from $0.2T$ to $3.1T$ in Fig.~\ref{subfig:cirleRMSE}. Constant $f_m$ results in large localization errors, especially when $T_m$ is an integer multiple of $T$. This occurs because most estimations of first half circle are usually from one jammer, and the second half attributed to another. This pattern also explains localization errors are lower while $T_m = 1.5T, 2.5T$. Once $T$ is accessed, synchronously taken \gls{aoa} estimations with $T$ can help minimize errors. Conversely, when $f_m$ exhibits randomness, making it inaccessible, its impact is typically reduced compared to a constant $f_m$.

    \section{Conclusion}\label{sec:con}
    In this paper, we \revise{}{have} thoroughly investigated the localization of power-modulated jammers using \gls{aoa} measurements obtained from a \gls{uav}. \revise{Precise localization was achieved even in complex scenarios involving multiple jammers where ideal \gls{aoa} resolution were not available. The proposed method, \gls{spgd}, demonstrated greater robustness compared to \gls{wlse} and \gls{lse} under all power modulation schemes, while maintaining lower computational complexity.}{Concerning complex practical scenarios involving multiple jammers and non-ideal \gls{aoa} resolution, we have proposed a novel \gls{spgd} solution for precise localization. It has been demonstrated superior to  \gls{wlse} and \gls{lse} under all power modulation schemes, regarding both robustness and computational complexity.}

    \section*{Acknowledgment}
	This work is supported partly by the German Federal Ministry of Education and Research within the project Open6GHub (16KISK003K/16KISK004), partly by the European Commission within the Horizon Europe project Hexa-X-II (101095759). B. Han (bin.han@rptu.de) is the corresponding author.
    
\bibliographystyle{IEEEtran}
\bibliography{references}

\begin{thebibliography}{10}
\providecommand{\url}[1]{#1}
\csname url@samestyle\endcsname
\providecommand{\newblock}{\relax}
\providecommand{\bibinfo}[2]{#2}
\providecommand{\BIBentrySTDinterwordspacing}{\spaceskip=0pt\relax}
\providecommand{\BIBentryALTinterwordstretchfactor}{4}
\providecommand{\BIBentryALTinterwordspacing}{\spaceskip=\fontdimen2\font plus
\BIBentryALTinterwordstretchfactor\fontdimen3\font minus \fontdimen4\font\relax}
\providecommand{\BIBforeignlanguage}[2]{{%
\expandafter\ifx\csname l@#1\endcsname\relax
\typeout{** WARNING: IEEEtran.bst: No hyphenation pattern has been}%
\typeout{** loaded for the language `#1'. Using the pattern for}%
\typeout{** the default language instead.}%
\else
\language=\csname l@#1\endcsname
\fi
#2}}
\providecommand{\BIBdecl}{\relax}
\BIBdecl

\bibitem{TJGJsecureUAV2019}
J.~Tang, G.~Chen, and J.~P. Coon, ``{Secrecy Performance Analysis of Wireless Communications in the Presence of {UAV} Jammer and Randomly Located {UAV} Eavesdroppers},'' \emph{IEEE TIFS}, vol.~14, no.~11, pp. 3026--3041, 2019.

\bibitem{ZYGMsecureUAV2018}
Y.~Zhu, G.~Zheng, and M.~Fitch, ``{Secrecy Rate Analysis of {UAV}-Enabled mmWave Networks Using Matérn Hardcore Point Processes},'' \emph{IEEE JSAC}, vol.~36, no.~7, pp. 1397--1409, 2018.

\bibitem{WQWRjamminguav2019}
Q.~Wu, W.~Mei, and R.~Zhang, ``{Safeguarding Wireless Network with {UAVs}: A Physical Layer Security Perspective},'' \emph{IEEE Wirel. Commun.}, vol.~26, no.~5, pp. 12--18, 2019.

\bibitem{QXJTpdr2015}
W.~Qipin, W.~Xianglin, F.~Jianhua \emph{et~al.}, ``{A step further of {PDR}-based jammer localization through dynamic power adaptation},'' in \emph{IEEE WiCOM 2015}, 2015, pp. 1--6.

\bibitem{WXWSurvJLoc2016}
X.~Wei, Q.~Wang, T.~Wang \emph{et~al.}, ``{Jammer Localization in Multi-Hop Wireless Network: A Comprehensive Survey},'' \emph{IEEE COMST}, vol.~19, no.~2, pp. 765--799, 2017.

\bibitem{LZHWYLSQ2012}
Z.~Liu, H.~Liu, W.~Xu \emph{et~al.}, ``{Exploiting Jamming-Caused Neighbor Changes for Jammer Localization},'' \emph{IEEE TPDS}, vol.~23, no.~3, pp. 547--555, 2012.

\bibitem{HAAjRSSD2020}
A.~Heydari and M.~Aghabozorgi, ``{Joint RSSD/AOA Source Localization: Bias Analysis and Asymptotically Efficient Estimator},'' \emph{Wirel. Pers. Commun.}, vol. 114, pp. 2643--2661, 2020.

\bibitem{QHLXsourceloc2020}
H.~Qi, L.~Mo, and X.~Wu, ``{SDP Relaxation Methods for RSS/AOA-Based Localization in Sensor Networks},'' \emph{IEEE Access}, vol.~8, pp. 55\,113--55\,124, 2020.

\bibitem{SBYAOAloc2023}
B.~Shi, Y.~Li, G.~Wu \emph{et~al.}, ``{Low-Complexity Three-Dimensional AOA-Cross Geometric Center Localization Methods via Multi-UAV Network},'' \emph{Drones}, vol.~7, no.~5, 2023.

\bibitem{TDGjammingpow2022}
P.~Tedeschi, G.~Oligeri, and R.~Di~Pietro, ``Localization of a power-modulated jammer,'' \emph{Sensors}, vol.~22, no.~2, 2022.

\bibitem{OAMjammDoA2020}
A.~Osman, M.~M.~E. Moussa, M.~Tamazin \emph{et~al.}, ``{DOA Elevation and Azimuth Angles Estimation of GPS Jamming Signals Using Fast Orthogonal Search},'' \emph{IEEE TAES}, vol.~56, no.~5, pp. 3812--3821, 2020.

\bibitem{MUSI1986}
R.~Schmidt, ``{Multiple emitter location and signal parameter estimation},'' \emph{IEEE TAP}, vol.~34, no.~3, pp. 276--280, 1986.

\bibitem{FOSDOAKM1988}
M.~J. Korenberg, ``{Identifying nonlinear difference equation and functional expansion representations: the fast orthogonal algorithm},'' \emph{Ann. Biomed. Eng.}, vol.~16, pp. 123--142, 1988.

\bibitem{ESDOARRK1989}
R.~Roy and T.~Kailath, ``{ESPRIT-estimation of signal parameters via rotational invariance techniques},'' \emph{IEEE TASSP}, vol.~37, no.~7, pp. 984--995, 1989.

\bibitem{LSEW2021}
F.~Watanabe, ``{Wireless Sensor Network Localization Using AoA Measurements With Two-Step Error Variance-Weighted Least Squares},'' \emph{IEEE Access}, vol.~9, pp. 10\,820--10\,828, 2021.

\bibitem{GDzbs2023}
Z.~Fang, B.~Han, and H.~D. Schotten, ``{A Reliable and Resilient Framework for Multi-UAV Mutual Localization},'' in \emph{IEEE VTC2023-Fall}, 2023, pp. 1--7.

\end{thebibliography}

\end{document}